\documentclass[iop]{emulateapj}

\usepackage{graphicx}
\usepackage{txfonts}
\usepackage{color}
\usepackage{natbib}
\bibpunct{(}{)}{;}{a}{}{,} %Astronomy & Astrophysics, ApJ

\newcommand{\hii}{H\,{\sc ii}~}

\shortauthors{Vicente, Bern\'e, Tielens et al.}  
\shorttitle{PAHs in the proplyd HST10}

\begin{document}

\title{PAH emission in the proplyd HST10: what is the mechanism behind photoevaporation? }
\author{S. Vicente\altaffilmark{1}, O. Bern\'e\altaffilmark{2,3}, A. G. G. M. Tielens\altaffilmark{4}, N. Hu\'elamo\altaffilmark{5},   E. Pantin\altaffilmark{6}, I. Kamp\altaffilmark{1}, A. Carmona\altaffilmark{7}}
%
%
%\email{olivier.berne@irap.omp.eu}

\altaffiltext{1}{Kapteyn Astronomical Institute, Postbus 800, 9700 AV, Groningen, The Netherlands}
\altaffiltext{2}{Universit\'e de Toulouse; UPS-OMP; IRAP;  Toulouse, France}
\altaffiltext{3}{CNRS; IRAP; 9 Av. colonel Roche, BP 44346, F-31028 Toulouse cedex 4, France}
\altaffiltext{4}{Leiden Observatory, Leiden University, Niels Bohrweg 2, NL-2333 CA Leiden, The Netherlands} 
\altaffiltext{5}{CAB (INTA-CSIC), LAEFF, P.O. Box 78, 28691 Villanueva de la Ca\~nada, Madrid, Spain}
\altaffiltext{6}{Laboratoire AIM, CEA/DSM - CNRS - Universit\'e Paris Diderot, IRFU/SAP, 91191 sur Yvette, France}
\altaffiltext{7}{UJF-Grenoble 1/CNRS-INSU, Institut de Plan\'etologie et d'Astrophysique de Grenoble (IPAG) UMR 5274, Grenoble, F-38041, France}

\begin{abstract}

Proplyds are photodissociation region (PDR)-like cometary cocoons around young stars which are thought to originate through photo-evaporation of the central protoplanetary disk by external UV radiation from the nearby OB stars.
%Proplyds are young stellar objects surrounded by protoplanetary disks and cometary cocoons which are interpreted  by photo-evaporation models as a photodissociation region (PDR) produced by the external UV radiation of nearby OB starts. 
This letter presents spatially resolved mid-infrared imaging and spectroscopy of the proplyd HST10 obtained with the VLT/VISIR instrument. These observations allow us to detect Polycyclic Aromatic Hydrocarbons (PAH) emission in the proplyd photodissociation region and to study the general properties of PAHs in proplyds for the first time. We find that PAHs in HST10 are mostly neutral and at least 50 times less abundant than typical values found for the diffuse ISM or the nearby Orion Bar. With such a low PAH abundance, photoelectric heating is significantly reduced. If this low abundance pertains also to the original disk material, gas heating rates could be too low to efficiently drive photoevaporation unless other processes can be identified. Alternatively, the model behind the formation of proplyds as evaporating disks may have to be revised.
 
\end{abstract}


\section{Introduction}

%which are being photoevaporated by the external UV radiation of nearby massive stars. Models of proplyds consider the photoelectric effect on Polycyclic Aromatic Hydrocarbons (PAHs) and small dust grains as the main mechanism heating the gas at the disk surface and driving the neutral evaporation flow and mass-loss in proplyds. The Far-UV photons dissociate molecules and create a phodissociation region (PDR)  which extends from the disk surface to the proplyd ionization front. However, PAHs have never been directly observed in the PDR of a proplyd. 

Most low mass stars are born in transient OB associations \citep[e.g.,][]{Lada2003}, and there is evidence that our own young Solar System evolved near massive stars (e.g., \citealt{Hester2004}). Externally illuminated protoplanetary disks or \emph{proplyds} \citep{O'Dell1993} are young stellar objects (YSOs) surrounded by Solar System-sized protoplanetary disks and found embedded within or near a \hii region. In these extreme environments the disks are exposed to intense UV radiation fields and stellar winds from the OB stars, dynamical encounters with sibling stars and supernovae, on timescales of planetary system formation and early evolution.  Hence, the study of their properties may bring key constraints on the understanding of the general mechanism of planet formation and the origins of our Solar System.

Proplyd morphology has been explained by models of evaporating flows in externally illuminated disks or globules \citep[e.g.,][]{Henney1996, Sutherland1997, Johnstone1998, Henney1998, Storzer1998, Storzer1999, Richling1998, Richling2000,Vasconcelos2011} as a FUV (6~\mbox{eV}$\le$~E~$<$~13.6~\mbox{eV}) heated photodissociation region (PDR) encased within a EUV (E~$\ge$~13.6~eV) ionized hydrogen cocoon with a ``head-tail" shape.
FUV photons of nearby OB stars penetrate deep into the disk and, at relative high column densities, dissociate hydrogen molecules, ionize carbon, and heat the gas to T$\sim$400-4000~K, forming the PDR. The resulting thermal pressure drives an expanding hydrodynamical flow of neutral material escaping from the disk surface with velocities of \mbox{1-3~kms$^{-1}$}. The \mbox{mass-loss} rate generated by this photo-evaporation wind determines the lifetime of the gaseous disk and hence the timescale for the formation of giant planets. At some distance from the disk, the EUV photons ionize the neutral wind and form an ionization front (IF) with T$\sim$10$^4$~K. The observed tails result from the diffuse UV radiation field (produced through the recombination of surrounding nebular gas) which drives an evaporation flow on the shadowed side of the disk. 
 These models consider the FUV  photoelectric effect on small dust grains and Polycyclic Aromatic Hydrocarbons (PAHs, \citealt{Joblin2011})  as the main gas-heating mechanism at the disk surface. %and driving the neutral evaporation flow and mass-loss in proplyds. 
Recent observations (Okada et al., \emph{submitted}) show that PAHs play a major role in the heating process, and theoretical studies \citep{Kamp2004} confirm this also for protoplanetary disks.
The proplyd models mentioned above are adapted from classical PDR models (see \citealt{Hollenbach1997} for a review) which  consider standard interstellar medium (ISM) PAH abundances. However, up to now, there is no  observational abundance determination for proplyds. 

%-------------------------------------------------------------------------------------------
%                                                  FIGURE 1
%-------------------------------------------------------------------------------------------
 \begin{figure*}[!th]
\includegraphics[width=\textwidth]{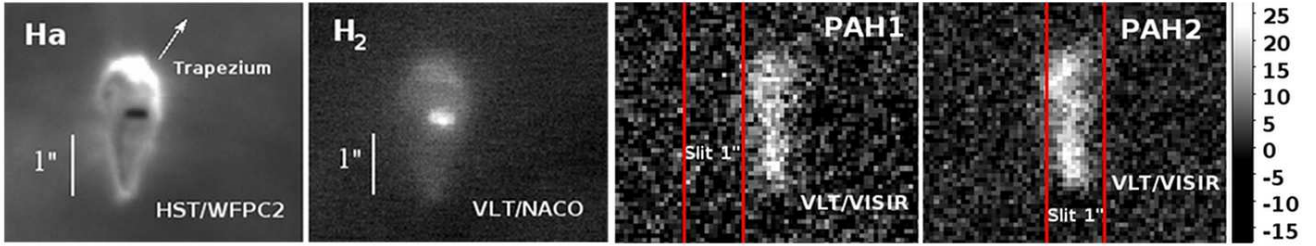}

\caption{HST10 images (from left to right): H$\alpha$ (0.656 $\mu$m)  from HST/WFPC2 (45.6 mas/pix; HST archive), H$_2$~(2.12 $\mu$m)  from VLT/NACO (27 mas/pix; Vicente et al., \emph{in prep.}), and PAH1 (8.6~$\mu$m) and PAH2 (11.25 $\mu$m) from VLT/VISIR (75 mas/pix; this work). The VISIR images show the 1\farcs0 slit position for  the 8.6 and 11.4 $\mu$m spectral settings respectively with the intensity scale given in mJy/arcsec$^2$. The arrow indicates the direction towards \mbox{$\theta ^1$ Ori~C}, the main ionizing O-star of the Trapezium cluster. North is up and east to the left.}

\label{fig1}
\end{figure*}
%------------------------------ end of Fig. 1 ---------------------------------

%Mid-IR observations of proplyds in Orion \citep{Hayward1997, Robberto1999, Robberto2002, Robberto2005} detected mid-IR excesses in accordance to the presence of circumstellar disks and, for a few of them, imaging with the Gemini-south (8m) was able to spatially resolve extended emission at 11.7 $\mu$m from tails, Herbig-Haro jets and bow shocks \citep{Smith2005}. \cite{Shuping2006}  detected strong silicate emission for seven (out of a sample of eight) Orion proplyds in \mbox{8-13~$\mu$m} low-resolution spectra (R$\sim$100) obtained with Keck. Extended PAH emission associated to three Orion proplyds was observed during the science verification of VISIR (PI, N.~Hu\'elamo), the mid-infrared spectrometer and imager at the VLT.  However, there are no other studies attesting for the presence of PAHs in the disks or PDRs of proplyds .

HST10 (182-413; \citealt{O'Dell1994}) is a teardrop-shaped proplyd (1\farcs2 $\times$ 2\farcs6)  containing a prominent nearly edge-on disk (i$\sim$80$^{\circ}$, PA$\sim$86$^{\circ}$) visible as a dark silhouette in H$\alpha$, optical ionized species and the continuum, but glowing in [OI] $\lambda$6300 and in the 2.12~$\mu$m~ro-vibrational line of H$_2$ (\citealt{Bally2000, Chen1998}, Vicente et al., \emph{in prep.})  This object is located at a projected distance of 56" from $\theta^1$ Ori C to the SE, the main ionizing star of the Trapezium Cluster (414$\pm$7~pc, \citealt{Menten2007}).  In this letter we present spatially resolved VLT/VISIR observations of PAH emission in HST10.  We determine the PAH abundance in the neutral flow and discuss the implications for photo-evaporation models.

\section{Observations and data reduction}

Mid-IR imaging (75~mas/pix) and low-resolution spectroscopy (R$\sim$200-350, 127~mas/pix) of the proplyd HST10 were obtained on the 13th December 2005 with VISIR \citep{Lagage2004}, under good ambient conditions  (seeing at 0.5~$\mu$m between 0\farcs6 and 1$''$). A spectro-photometric standard (HD\,35536) of similar airmass was observed immediately before and after the science target to allow for telluric absorption correction and calibration. The data were reduced with a VISIR customary pipeline based on IDL scripts \citep{Pantin2005} and corrected for the background using a multi-resolution inpainting scheme \citep{Pantin2010}. %The nebular back/foreground nearby the proplyd shows considerable spatial structure in all mid-IR wavelengths with HST10 laying on a ridge of emission at PA$\sim$130$\degr$. 
The images were collected with filters PAH1 (8.59$\pm$0.21~$\mu$m), SIV (10.49$\pm$0.08~$\mu$m), and PAH2 (11.25$\pm$0.3~$\mu$m) for a total exposure time (on source) of 30min and using the parallel chopping/nodding jitter mode with a chopping throw of  7$''$. The spatial resolution of the final images ($19" \times 19"$), measured in the only point source in the field, was 0\farcs35 for PAH1 and 0\farcs34 for PAH2, the latter being diffraction limited and reflecting the improvement in the seeing\footnote{The diffraction limit at the VLT (8.2m) varies from  0\farcs26 to  0\farcs34 in the range 8.6 -- 11.25$\mu$m. As the size of a UT mirror is comparable to the turbulence outer scale, VISIR data are already diffraction limited for optical seeing below 0\farcs6.}. HST10 is spatially resolved in the PAH1 and PAH2 images (0\farcs97$\times$ 2\farcs5), but not detected in SIV after nebular subtraction. No central star inside the disk is visible in any of the VISIR images (Fig.~1). 

The $N$-band spectroscopy consisted of three settings centered at 8.5, 9.8 and 11.4~$\mu$m (1h integration time) chosen for covering the 10$\mu$m silicate feature and PAH emission bands at 8.6 and 11.25~$\mu$m. 
The 1$''$~slit was placed along the proplyd head-tail, in the NS orientation, and the chop/nod was performed along the slit with a throw of~8$''$. The 1D-spectra were extracted by integrating the flux for each wavelength over the spatial extension of the proplyd ($\sim$17-19 pix) in the 2D background subtracted spectra. They show spatially resolved PAH emission at 11.25~$\mu$m (Fig.~2) but no detection above the noise level at 8.6~$\mu$m.  The slit position, estimated using the RA and DEC values from the headers, is overlaid on top of the VISIR images in Fig.~1 and have an uncertainty of 0\farcs2-0\farcs3, corresponding to the accuracy  of (small)  relative offsets with the VLT. The RA offset of the slit at 8.6$\mu$m, relative to the proplyd nominal position in \cite{O'Dell1994}, is 0\farcs6-0\farcs8 to the east which leaves the proplyd out of the slit.  We believe this was due to errors occurring during the blind offsetting performed to obtain the HST10 spectra.  For the 11.4~$\mu$m setting  the proplyd falls almost entirely inside the slit (see Fig.~1).

%-------------------------------------------------------------------------------------------
%                                                  FIGURE 2
%-------------------------------------------------------------------------------------------

 \begin{figure}[!ht]
\flushright
\includegraphics[width=\hsize]{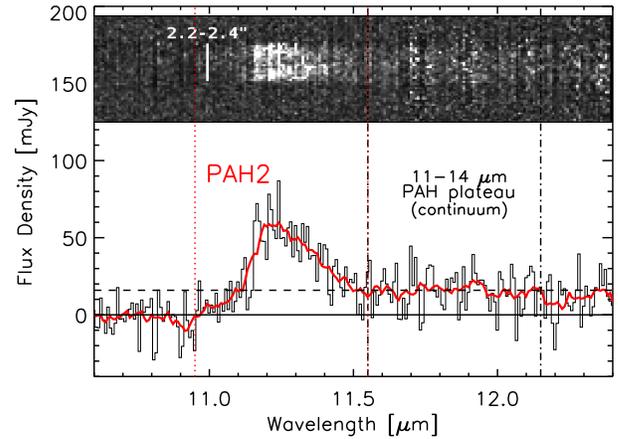}

\caption{VLT/VISIR low-resolution spectrum (R$\sim$200) of the proplyd HST10 showing the broad PAH band at 11.25 $\mu$m, and extracted from the 2D spectrum on the top. The PAH2 filter range is indicated by the vertical dotted lines. A flux of 16 mJy for the PAH plateau was determined as the mean flux in an interval of the same length as the PAH2 filter and  is represented by the horizontal dashed line. Over-plotted is the spectrum smoothed with a boxcar = 8. } 

\label{fig2}
\end{figure}

%------------------------------ end of Fig. 2 ---------------------------------
%-------------------------------------------------------------------------------------------
%                                                  FIGURE 3
%-------------------------------------------------------------------------------------------

 \begin{figure*}[!th]
 \includegraphics[width=\hsize]{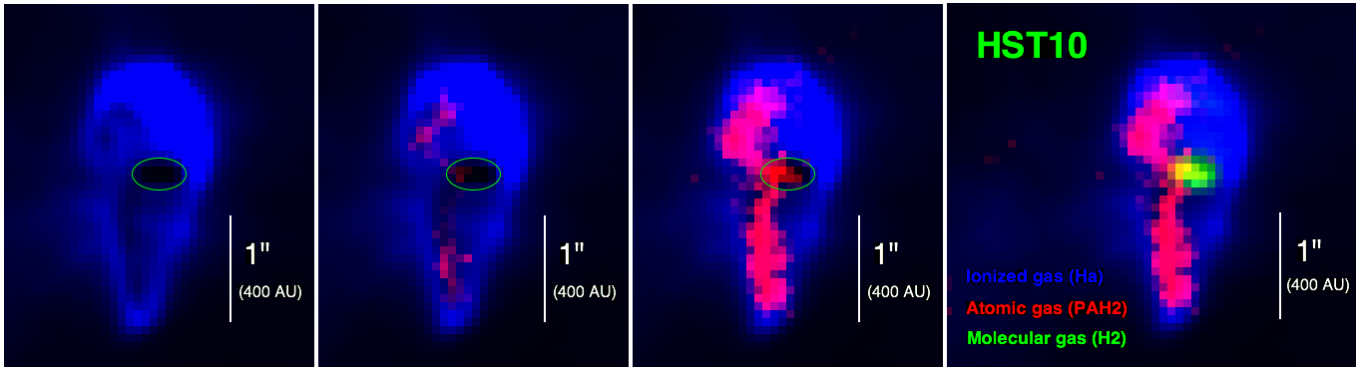}
\centering
\caption{Color composite image of the proplyd HST10 applying a square-root stretch in intensities for better contrast, and showing the different regions traced by the different wavelengths: H$\alpha$ tracing the ionized gas at the ionization front (blue), H$_2$ the molecular gas at the surface of the nearly edge-on protoplanetary disk (green), and the PAHs emission the atomic gas within the PDR (red). From left to right we increase progressively the intensity of the PAH2 emission and find it to be coincident with the more extincted  or less emitting regions in the H$\alpha$ optical image. The silhouette disk is indicated by the green ellipse. }
\label{fig3}

\end{figure*}

%------------------------------ end of Fig. 3 -------------------------------------------------------

The H$_2$ 2.12~$\mu$m image (27~mas/pix) was collected with the VLT adaptive optics instrument NACO under program ID~076.C-0874 (PI, S. Vicente). This data is described and analyzed in a forthcoming paper (Vicente et al., \emph{in prep.}). The H$\alpha$~0.656$\mu$m  image (45.6 mas/pix) from the Hubble Space Telescope instrument WFPC2 was retrieved from the ESO archive (program GO~6603, J.~Bally). The optical and \mbox{near-IR} images were rebinned to the same pixel size of the VISIR images and aligned with sub-pixel precision (0.01 pix) with IMALIGN/IRAF. The NACO image was used as reference because it contained several point sources in the field, contrary to the VISIR and HST images for which only one to two stars could be found. Although there is a perfect overlap of the centroids of the stars in the RGB image, the estimated error in the alignment is  $\pm$\,1pix   (or $\pm$\,75~mas) from comparison of the position of the silhouette disk  seen in the H${\alpha}$ image to the H$_2$ disk emission seen in the NACO image (Fig.~3). This error results from the different Point Spread Functions (PSFs)  observed at the different wavelengths (0\farcs06 for H${\alpha}$, 0\farcs08 for H$_2$, and 0\farcs35 for PAHs). 

\section{Observational results}\label{results}

Fig.~1~shows the optical, near-IR and mid-IR images of HST10 tracing the different key elements of its morphology. The optical H$\alpha$ image traces the ionized gas at the ionization front with its peak surface brightness facing the Trapezium stars. This is a clear evidence that  $\theta^1$ Ori C  is the main source of UV radiation driving the photo-ionization and evaporation in HST10. The H$_2$~2.12~$\mu$m emission traces the molecular gas at the disk surface. This was first discovered in a HST/NICMOS image \citep{Chen1998} and more recently confirmed with adaptive optics ground-based imaging (Vicente et al., \emph{in prep.}). 
The VISIR mid-IR images reported here show emission associated to the proplyd in both filters PAH1 (8.6~$\mu$m) and PAH2 (11.25~$\mu$m). The latter detection is also confirmed with spectroscopic observations (Fig.~2) and can be attributed to (solo) C-H out-of-plane bending mode in PAH molecules with long straight edges.  No continuum is observed below 11$\mu$m, but beyond 11.3 $\mu$m we see in Fig.~2 the 11-14 $\mu$m PAH plateau resulting from the blend of C-H out-of-plane bending modes (e.g. arising from PAH clusters). On the basis of the similarity of the morphologies in the PAH1 and PAH2 images, and since the PAH1 filter is also situated on top of a C-H vibration (in-plane bending mode), we conclude the PAH1 emission is also due to PAHs. 
The total density flux  integrated over the full proplyd extension (rectangular aperture of $25 \times 45$~pix) in the PAH1 image is 26.6$\pm$0.5~mJy (3$\sigma$) when considering only the photon noise. Given that the uncertainty in the conversion factor used for calibrating the images is typically 10\%, we obtain 26.6$\pm$2.6~mJy. The integrated flux of the 11.25~$\mu$m feature in the spectrum, divided by the passband of the PAH2 imaging filter (10.95$-$11.55~$\mu$m), is 29.4$\pm$12.9~mJy (1$\sigma$) which is consistent,  within the error, with the extracted photometry in the PAH2 image, 38.8$\pm$3.8~mJy.
The smaller value obtained from the spectrum with respect to imaging is likely due to slit losses which were worsen by pointing inaccuracies from the blind offsetting. 
The 11.25~$\mu$m emission observed in the 2D-spectrum in Fig.~2 is extended over $\sim$17-19 pixels or 2\farcs2-2\farcs4 (127~mas/pix), similar to the head-to-tail size of HST10 in the PAH images, and is about 50 mJy at the  peak position. The maximum surface brightness in the PAH1 and PAH2 filters ($I_{8.6}$ and $I_{11.25}$) were derived by taking the average of  the four brightest pixels at the proplyd head in each VISIR image. These values are given in Table 1.

\section{Properties of PAHs in HST10}

\subsection{Spatial distribution and ionization}\label{ionization}

In PDRs, PAHs emit mostly in the neutral gas  where the majority of  molecules are dissociated by FUV photons (e.g. in the nearby Orion Bar, \citealt{Tielens1994}). Hence, PAHs are a tracer of the atomic gas (for instance in the NGC 7023 nebula they follow the far infrared emission of~C$^+$, \citealt{Joblin2010}), and in the particular case of proplyds,  PAHs will trace the morphology of the photo-evaporating flow.
This is well illustrated in Fig.~3 showing the ionized (H$\alpha$), neutral (PAHs) and molecular (H$_2$) gas components of the proplyd HST10. The PAH  emission is localized within the ionized envelope and at the disk surface from where, according to PDR models, the FUV-heated material evaporates generating the neutral wind and creating the proplyd PDR. Additionally, even thought the PAH1 and PAH2 emission show similar head-to-tail extent, their distribution is different. PAH1 emission at 8.6~$\mu$m is fainter ($\sim2 \times$ in surface brightness) and more homogeneously distributed within the cocoon, whereas the PAH2 emission at 11.25 $\mu$m is sharper and brighter on the opposite side to the direction of $\theta^1$~Ori~C,  and coincident with the areas showing less emission or more heavily extincted in the optical H$\alpha$ image (Fig.~3).  Considering that most of the optical extinction is caused by $\sim$0.1$\mu$m dust grains, and these are expected to be depleted in the neutral flow (with $A_\mathrm{V} = 0.1- 0.2$, \citealt{Henney1999}) due to grain growth and settling in the protoplanetary disk, the 11.25 $\mu$m emission is tracing the regions of higher density in atomic gas and possibly very small particles of dust.  Additionally, while the 11.25~$\mu$m feature is dominant for neutral PAH molecules, the 8.6~$\mu$m feature is stronger for ionized PAHs mostly present in high UV irradiated regions   (\citealt{Joblin2011}, and references there in).
Hence, the difference in spatial distribution observed in the two filters may reflect a charge effect associated to the proplyd location in the foreground of the nebula, that is, in between $\theta^1$ Ori C and the observer. Positively ionized PAHs are expected to be more abundant on the irradiated side of the proplyd opposite to us, whereas  the bulk of the PAHs reservoir, as seen in the proplyd ``shadowed" side,  seem to be in the form of neutral molecules. The  $I_{8.6}/I_{11.25}$ band ratio, at the position where we extracted brightnesses in HST10, is of the order of 0.5, in accordance to the astronomical template of \citet{Pilleri2012} for which a value around 0.4 is found for neutral PAHs and 1.45 for PAH$^+$.
For high radiation fields, PAHs can be neutral if the density of the gas in the flow is high ($\sim10^6$ cm$^{-3}$,  \citealt{Tielens2005}), allowing for efficient recombination of PAH cations with slow electrons. As we will see, this is most likely the case (Sect.~\ref{abundance}). 
Finally, we note that there is no PAH emission beyond the ionization front, suggesting that they are largely destroyed beyond this point as seen in the Orion Bar \citep{Giard1994}.

%------------------------------------------------------------------------------
%                                             TABLE 1
%------------------------------------------------------------------------------
\begin{table}[!thb]

\caption[]{
Parameters of NGC\,7023 and HST10 used to derive the PAH abundance in equation~1}
\label{table:1}
\bigskip
\centering
\begin{tabular}{lll}
\hline\hline
\noalign{\smallskip}
Parameter &  &  Reference \\
\noalign{\smallskip}
\hline
\noalign{\smallskip}
\multicolumn{3}{c}{NGC 7023}\\
\noalign{\smallskip}
\hline
\noalign{\smallskip}
$I_{8.6}$  &   1306 $\pm$ 5  MJy\,sr$^{-1}$ &  this paper \\
\smallskip
$I_{11.25}$ &   2723 $\pm$ 6 MJy\,sr$^{-1}$ & this paper \\ 
\smallskip 
$G_0$ &   $2.6 \times 10^3$  & \citealt{Joblin2010} \\
\smallskip
N$_H$   & $1 \times 10^{22}$ cm$^{-2}$ & \citealt{Joblin2010}  \\
\smallskip
$f_C^{PAH}$  & $7 \times 10^{-2}$ & \citealt{Berne2012}  \\
\noalign{\smallskip}
\hline
\noalign{\smallskip}
\multicolumn{3}{c}{HST 10}\\
\noalign{\smallskip}
\hline
\noalign{\smallskip}
$I_{8.6}$  &   720 $\pm$ 111  MJy\,sr$^{-1}$ &  this paper \\
\smallskip
$I_{11.25}$ &   1466 $\pm$ 111 MJy\,sr$^{-1}$ & this paper \\ 
\smallskip
$I_{8.6}/I_{11.25}$  &  0.5  & this paper \\
\smallskip 
$G_0$ &   $2.4 \times 10^5$  & \citealt{Storzer1999} \\
\smallskip
N$_H$  & $5.5 \times 10^{21}$ cm$^{-2}$ & \citealt{Storzer1999}  \\
\smallskip
$f_C^{PAH}$  & $8 \times 10^{-4}$ & this paper  \\
\hline\hline
\end{tabular}
\end{table}
%------------------- END of Table 1 -----------------------------------

\subsection{PAH abundance}\label{abundance}

Proplyd models consider disk photoevaporation to be due mainly to efficient heating of the gas by energetic photo-electrons provided by small grains and PAH molecules. However, these models  assume  PAHs in proplyds are as abundant as in the interstellar medium (ISM), an assumption which has not yet been verified with observations. 
PAHs have been detected in, at most, 15\% of the observed disks around isolated TTauri stars and they are under-abundant by a factor of 25 when compared to the ISM \citep{Geers2007, Oliveira2010}. The VISIR images of HST10 presented in this paper can be used to estimate the abundance of PAHs in a proplyd PDR for the first time.

The \mbox{mid-IR} PAH emission  at a given wavelength $I_{\lambda}$ is proportional to the number of carbon atoms locked in PAHs  in the line of sight (e.g. \citealt{Joblin2010}), and on the intensity of the UV radiation field, $G_0$\footnote{expressed in terms of the Habing field which corresponds to an integrated intensity between 91.2 and 240nm of $1.6 \times 10^{-3}$ ergs\,cm$^{-2}$\,s$^{-1}$ \citep{Habing1968}.}. Hence, we can write 
\begin{equation}
I_\lambda=\epsilon_{\lambda}~f_C^{PAH}~\frac{[C]}{[H]}~N_H~G_0,
\label{pahem}
\end{equation}
where $f_C^{PAH}$ is the fraction of elemental carbon locked in PAHs, $N_H$ is the column density of hydrogen atoms in the line of sight,  ${[C]}/{[H]}$ is the abundance of carbon relative to hydrogen atoms ($1.6 \times 10^{-4}$), and $\epsilon_{\lambda}$ is the PAH emissivity at the given wavelength. The latter parameter, $\epsilon_{\lambda}$ can be derived for sources where all the other parameters given in Eq.~\ref{pahem} can be determined independently as in the case of  the reflection nebula NGC\,7023. The values adopted for $f_C^{PAH}$, $N_H$ and $G_0$ were taken from the literature (Table~1) while the intensities $I_{8.6}$ and $I_{11.25}$ for  NGC\,7023 were measured directly in the \emph{Spitzer} IRS spectrum \citep{Pilleri2012}. This yields for the PAH emissivities $\epsilon_{8.6}=4.5\times10^{-18}$
 and $\epsilon_{11.25}=9.4\times10^{-18}$~MJy\,sr$^{-1}$cm$^{-2} G_0^{-1}$ which, when inserted in Eq. \ref{pahem} combined with the parameters  in Table~1, give the PAH abundance in the PDR of HST10, $f_C^{PAH}$, if a good estimate of the column density of atomic gas in the line of sight $N_H$ is provided. 
 % , at 8.6 and 11.25 $\mu$m respectively.  These values can then be used in Eq. \ref{pahem}, together with the parameters  in Table~1, to determine the PAH abundance $f_C^{PAH}$ in the PDR of HST10, provided we have a good estimate of the column density of atomic gas in the photo-evaporation flow on the line of sight, $N_H$.
This parameter can be estimated from the electron density at the ionization front which can be measured with a fair accuracy. \cite{Bally1998} obtain a value of $n_e=8 \times 10^4$~cm$^{-3}$ from the H$\alpha$ surface brightness, while \cite{Storzer1999} find  $n_e=1.1 \times 10^5$~cm$^{-3}$,  when correcting for the extinction to the Orion Nebula. Assuming pressure equilibrium at the ionization front, the density of H atoms in the neutral flow $n_H$, must be of the order of $10\times n_e$ or $n_H\sim10^6$~cm$^{-3}$. For a head width of $\sim 6\times10^{15}$~cm, measured in the PAH2 image, and assuming this value for the neutral flow lenght along the line of sight (symmetry in HST10), we obtain $N_H\sim6\times10^{21}$~cm$^{-2}$. From models and using the same parameters, \cite{Storzer1999} find $N_H=5.5\times10^{21}$~cm$^{-2}$. Adopting their value for $N_H$, and using the brightnesses  $I_{8.6}$ and $I_{11.25}$ (Table~1) measured in the VISIR PAH1 and PAH2 images, we find similar PAH abundances in HST10 of $f_C^{PAH}=8.1\times10^{-4}$ and $f_C^{PAH}=8.0\times10^{-4}$.
%Note that the two results derived independently using two different images at two different wavelengths match very well. 
These values are nevertheless extremely low: they correspond to an abundance of PAHs 90 times lower  than in NGC\,7023 \citep{Berne2012}, or about 50 times less than the values found in the Orion Bar or in the diffuse ISM \citep{Tielens2005}. And, since the brightnesses $I_\lambda$ have been measured for the brightest pixels at the proplyd head (Sect.~\ref{results}), our estimation gives the maximum abundance of PAHs in the evaporating flow toward $\theta^1$ Ori C. 

%PAHs are detected, and even brighter, in the proplyd tail meaning that the diffuse UV field is milder than the direct one, as expected.

%meaning that the diffuse UV field that is evaporating the gas on the shadowed part of the disk  and creating the proplyd tail is milder than the direct one, as expected.

\section{Discussion}

As other studies of TTauri stars \citep{Geers2007,Oliveira2010} we find PAHs to be under-abundant in the PDR of the proplyd HST10, by a factor of 50 or more relative to the diffuse ISM. 
The origin of this under-abundance cannot be readily explained, but some proposed hypotheses include clustering of PAHs followed by sedimentation inside the disk, destruction by FUV photons in the PDR, or destruction by X-rays emitted by the low-mass central star\footnote{The majority of stars in the Orion Nebula cluster are in the mass range of $0.1-1.5$~M$_\odot$ \citep{Hillenbrand1998}}.
Nevertheless, more important than the causes for PAH under-abundance are the implications this result has on our understanding of the physical processes shaping morphology and driving mass-loss in proplyds. 
Photo-evaporation of proplyds has been explained so far by the photo-electric heating of the gas which is known to have a much reduced photo-electric efficiency for grains larger than 100$\AA$ \citep{Tielens2005}.
Given the high column density $N_H=5.5\times10^{21}$~cm$^{-2}$ and low extinction $A_\mathrm{V} = 0.1- 0.2$ in the PDR of HST10, we expect these large grains to be depleted in the neutral flow. In fact, the photoelectric heating has it highest efficiency for PAHs (molecules of a few \AA) and the small end of the very small grains (up to a few tens of \AA). But according to \cite{Pilleri2012} (Fig.~6 in their paper), for high radiation fields as those found in proplyds ($> 10^4 G_0$), the very small grains are destroyed and evaporated into free-flying PAHs. Therefore, we do expect PAHs to be the main agents of the photo-electric heating in proplyds. 

 Assuming the disk surface has the same PAH abundance as the proplyd PDR (they are lifted from the disk surface by the evaporative wind),  the low fraction of PAHs found in this letter will have a profound impact on the disk surface gas temperature. And hence, the current hypothesis of an evaporating disk creating the proplyd PDR and morphology may have to be revised.
Two possibilities arise: 1) other gas heating mechanisms are relevant for disk evaporation,  such as collisional de-excitation of UV pumped H$_2$ and H$_2$ photo-dissociation followed by reformation on grain surfaces; 2) the PAH emission from the proplyd cocoon is associated to remnant atomic gas from the protostellar envelope or the surrounding nebula. 
%Nevertheless, the spatially resolved observations of HST10 in this paper demonstrate the disk must evaporate in order to create the observed morphology, and therefore the gas has to be hot. This raises the question of which is the main mechanism heating the gas. One possibility is collisional de-excitation of UV pumped H$_2$.
%For dense regions, as those expected at the surface of a protoplanetary disk, and depending on the temperature, the H$_2$ molecules  can compete with PAHs and dust for FUV photons and the heating efficiency by collisional de-excitation of H$_2$ can be comparable to that of the photoelectric effect \citep{Tielens2005}. Alternatively, heating by direct photoionization of the gas by X-rays from the central star could also be relevant. 
By combining detailed modeling with upcoming Herschel (Bern\'e et al., \emph{in prep.}) and VLT data (Vicente et al., \emph{in prep.}),  the heating-cooling mechanisms at the disk surface can be assessed allowing to test each one of the possible scenarios creating the puzzling proplyd morphology.

% to test the origin of disk evaporation (and hence \mbox{mass-loss}), and puzzling proplyd morphology. }

%In addition, a combined analysis of PAH properties (\mbox{mid-IR}) and far-IR cooling lines will be crucial to understand in better terms the role of PAHs  -- and their evolution --  in the heating of the gas, in both disks and the ISM.

%--------------------------------------------------------------

%\section{Conclusions}
%
%In this paper we derive the properties and abundance of PAHs in the Orion proplyd HST10 using spatially resolved observations from the VLT/VISIR instrument and empirical PAH emissivities measured in the NGC 7023 nebula. The results show that PAHs in the PDR are mostly neutral and depleted by a factor of at least 50 ($f_C^{PAH}=8.0\times10^{-4}$) relative to the diffuse ISM. These results are not in line with what is considered in PDR models used to explain gas heating and photoevaporation in proplyds. 
%
%\begin{acknowledgements}

%\end{acknowledgements}

%\bibliographystyle{apj}
%\bibliography{biblio.bib}

\end{document}